\documentstyle[11pt,newpasp,twoside,epsfig]{article}
\def\edcomment#1{\iffalse\marginpar{\raggedright\sl#1\/}\else\relax\fi}
\reversemarginpar

\begin{document}
\title{The spectroscopic redshifts of SCUBA galaxies: implications for spheroid
formation}
 \author{Scott C.\ Chapman$^1$,
Andrew W. Blain$^1$, Rob J. Ivison$^2$, and Ian Smail$^3$.
}
\affil{$^1$ Caltech, Pasadena, CA, 91125, USA}
\affil{$^2$ Royal Observatory, Edinburgh EH9 3HJ, UK}
\affil{$^3$ University of Durham, Durham DH1 3LE, UK}

\begin{abstract}
We present spectroscopic identifications for a sample of 55
submillimeter(submm)-selected `SCUBA' 
galaxies, lying at redshifts $z=0.7$ to 3.7,
that were pinpointed in deep 1.4-GHz VLA radio maps.
We describe their properties, especially the presence of active galactic 
nuclei (AGN) in the sample, and discuss the connection of the 
SCUBA galaxies and the formation of spheroidal components of galaxies, which 
requires knowledge of their masses and the
timescales of their very luminous activity. 
For a subset of the galaxies, we show their disturbed and diverse 
{\it Hubble Space Telescope (HST)} optical morphologies. 
\end{abstract}

\section{Introduction}

SCUBA galaxies (Smail, Ivison \& Blain 1997; Barger et al.\ 1998;
Hughes et al.\ 1998; Eales et al.\ 1999)
are an important population (Blain et al.\ 1999, 2002),
but frustratingly difficult to mesh with semi-analytic model frameworks
(Guiderdoni et al.\ 1998; C.Lacey et al., in preparation).
They are both numerous, with a surface density about 10\% that of 
optically-selected Lyman-break galaxies (LBGs; Steidel et al., 1999, 2003)
at the relative depths of the current surveys, 
and luminous (possibly because of very high SFRs), with typical bolometric 
luminosities 
about 10 times greater than LBGs, adopting plausible spectral energy 
distributions (SEDs). The existing SCUBA galaxies produce most of the 
extragalactic submm background radiation intensity, and a
significant fraction of the background at far-infrared (FIR) wavelengths 
(Blain et al.\ 2002; Smail et al.\ 2002; Cowie et al.\ 2002).
They represent a population which is often difficult to detect at
optical wavelengths, with at least the fainter half clearly being 
missed in current optical cosmological surveys.
Unfortunately, the SCUBA galaxy population is notoriously difficult to study!
Until now, we have been able to gather almost nothing about their redshifts 
and morphologies as fundamental observable properties.
As a consequence, deriving properties
such as total mass and dust temperature, and finding both the fraction of 
power contributed by AGN and starbursts, and their 
connection to optically-selected  
star-forming LBGs, have been the topic of largely idle speculation
over the five years since their discovery.

The principal hurdle has always been identification at other wavelengths.
SCUBA/MAMBO surveys have large beam sizes (10"--15"), a situation that	
will remain true for upcoming single-antenna, wide-field instruments: Bolocam, 
SHARC-II, a bolometer camera on the APEX telescope and SCUBA2.
Candidate counterparts to SCUBA galaxies cannot therefore be identified 
unambiguously without
interferometry to pinpoint their positions. However, mm/submm 
interferometry is currently an arduous process, with
tens of hours of integration required to detect a single object at
the OVRO MMA or IRAM PdB interferometers.

As a consequence, the 20-cm radio emission from SCUBA galaxies 
has become an important, but {\it second best} 
surrogate with which to probe both the 
energy generation processes and morphology of submm galaxies
(Ivison et al.\ 1998; Smail et al.\ 2000; Barger, Cowie \& Richards 2000; 
Chapman et al.\ 2001, 2002a, 2003a; Ivison et al.\ 2002).
The problem remains that in order to use a radio wavelength as a surrogate 
for submm/FIR emission,
we would like a clear physical principle connecting the emission at the 
different wavelengths. We don't yet, but 
we do have a strong empirical connection: the far-IR--radio correlation 
(e.g., Helou et al.\ 1985), which has an RMS dispersion of only 
0.2\,dex over a large range in luminosity at low redshifts.
Radio identification of SCUBA galaxies has allowed their optical properties
to be explored in detail (Chapman et al.\ 2003a; Ivison et al.\ 2002).
A large range of optical properties is observed for the radio-selected 
SCUBA galaxies, with 
65\% fainter than $I>23.5$. They have red optical-IR colors with $I-K=3$ to 6 
and $<I-K>=4.3$, but they are not all extremely red objects (EROs).
Deep radio observations allow SCUBA galaxies to be pre-selected with an 
efficiency of about 40\%.

\section{Redshifts for SCUBA galaxies}

We have been able to secure redshifts for 55 radio-identified submm galaxies
through deep Keck/LRIS observations (Chapman et al.\ 2003b, 
S.C.~Chapman, in preparation.), 
and the sample is growing rapidly.\footnote{Data presented herein were 
obtained at the W.M. Keck Observatory, which is operated as a scientific 
partnership between Caltech, the 
University of California and NASA. 
The Observatory was made possible by the generous financial
support of the W.M. Keck Foundation.} 
The sources lie in several distinct fields: CFRS03hr, Lockman-Hole, HDF,
SSA13, CFRS14hr, Elais-N2, SSA22. The spectra for several new 
examples are shown in Fig.~1.
The redshifts allow, for the first time, accurate calculation of 
luminosities and dust temperatures for the SCUBA galaxy population.
We emphasize that obtaining redshifts is not easy, relying heavily on the
superb blue sensitivity of the new LRIS-B mutliobject spectrograph 
(Steidel et al.\ 2003), and
often have no detectable continuum emission with which to extract the 
spectrum. 
In addition the galaxies are hard to identify, typically 
being faint, messy, composite objects
in optical images. Because of their
small radio/optical offsets ($\sim0.5$") it is often difficult
to assess the best position at which to align the slit. 
Often we designated several slit positions on different masks for 
each target.
Our spectroscopic completeness is $\sim$50\% over the magnitude range 
of the sample from $I=22$ to 27.

While the issue of correctly identifying the submm galaxy is 
concerning, three (out of 3) CO detections with IRAM-PdB have already been 
made,
realizing an unequivocal confirmation of the redshifts (Neri et al.\ 2003),
as well as dynamical mass estimates $>10^{11}$M$_\odot$. 
One goal of the IRAM-PdB program is to obtain CO detections for a
statistical sample $\sim$30 of 
our submm galaxies, to understand the range in molecular gas properties.
We also note that 
beyond the radio positional identification, we are 
finding the correct {\it type} of object through optical spectroscopy. 
The identifications are at high-$z$,
have apparent SFRs of several 100\,M$_\odot$\,yr$^{-1}$, 
often show AGN features, and are 
very rare objects in the LBG distribution. We ask, what else can they be?

We further note that we believe we have spectroscopically identified 
a sample of blank-field SCUBA galaxy counterpart candidates without
radio detections, effectively through trial and error by targeting faint 
optical sources 
lying within the SCUBA beam using otherwise redundant spectrograph slits.
Observations of these identifications were tried
based on our cumulative experience of with the properties
of the radio-identified sample: often 
faint, distorted blue/red composite sources near the SCUBA beam 
centroid, are found to have Type-II AGN spectra and/or inferred star formation 
rates of order 100\,M$_\odot$\,yr$^{-1}$.
Again, CO detection will be the final arbiter concerning the validity of the
identifications.
These identifications overlap significantly in redshift 
with our radio-submm sample,
their radio {\it non}-detections implying colder dust temperatures.

In Fig.~2, we show the observed redshift distribution, and a toy model
for the radio and submm distributions 
derived from an evolving far-IR luminosity function (Chapman et al.\ 2003c, 
Lewis et al.\ 2003).
The model is very useful for understanding 
selection effects, and in particular the bias against the highest-redshift 
galaxies due to the requirement of a radio selection. This bias has a 
strong dependence on dust temperature ($T_{\rm d}$). The
sources missed by the radio are expected to fill the region lying between the
model radio and submm distributions, overlapping significantly in redshift
with the radio-identified 
sample if ther dust temperatures are in the cold to warm
regime ($T_{\rm d}<35$\,K).
The redshift distribution of a 
radio-selected QSO sample (Shaver et al.\ 1998) (which is unlikely to be
affected
by dust obscuration) is overplotted, suggesting a remarkable correspondence
with the submm galaxy population.

If our identifications for submm galaxies {\it without} radio detections
are correct, then this leaves only a $\sim20$\% tail of submm galaxies
which can lie above $z>4$. With a surface density of $\sim200$/deg$^2$, this
20\% tail is still considerably larger than the $z>4$ QSOs detected in the 
mm/submm (Carilli et al.\ 2001, Isaak et al.\ 2002).
Continued effort to identify these sources will be a worthy pursuit of the 
newest instruments and techniques.

\section{Spheroids in Formation?}

We have presented a plausible redshift distribution for submm galaxies, 
the most 
important ingredient for addressing whether they could be spheroids in 
formation.
Encouragingly, they lie at the correct redshifts to be proto-ellipticals or 
the forming bulges of spiral galaxies. 
(all the stellar population constraints point to most stars being formed 
at $z$=2--3). 
In particular, we have demonstrated that most submm galaxies do not lie at 
$z>4$, and thus most are not very high redshift
Population-III sources.
Quite surprisingly, even this small sample of spectroscopic identifications
reveals a redshift clustering signal, bolstering their association with 
massive halos (Blain et al.\ 2003). 

However, to assess what type of formation mechanism is at work
we must study the AGN/starburst contribution to the dust heating,
the timescales of their luminous phase, and compare their {\it HST} 
morphologies.
The AGN versus starburst issue is always difficult to address.
We have firstly from our restframe UV spectra, the possibility of
diagnosing the presence of an AGN through high ionization lines.
Indeed, we find signs of CIV, and other lines which cannot easily be
excited in a starburst, in approximately 50\% of our sources.
Our single near-IR spectrum thus far (Smail et al.\ 2003) 
reveals an OIII/H$\beta$ ratio
typical of Seyfert-2 galaxies as well as strong NeV emission.
X-ray detections of 7 submm galaxies in the Chandra Deep Field North were
presented in Alexander et al.\ (2003), while we have measured significant
X-ray flux from a further seven of our sample, together implying that 
$\sim$70\% of the radio-identified SCUBA galaxies have X-ray detections.
However, only 1/3 of these are too X-ray bright 
to be generated by star formation alone (scaling from Nandra et al.\ 2002
for LBGs).
Finally, $\sim$20\% of our sources appear to be unusually bright at radio 
wavelengths,
departing significantly from the far-IR--radio distribution for 
low-redshift starburst
galaxies.
Together, this suggests that AGN are prevalent in the sample
(as expected from the likely coeval evolution of the BH and bulge required 
to generate the tight correlation between their masses; Magorrian et al.\
1998), 
but not necessarily dominant in the
bolometric energy released by SCUBA galaxies.

The duration of the very luminous activity of the SCUBA galaxies are 
important: are their timescales 10\,Myr or 1\,Gyr?
If the former, then several bursts would be required to form a massive 
elliptical galaxy. In the spectra with the highest signal-to-noise 
ratios, stellar and interstellar
features have been observed. These can be fitted using starburst synthesis 
models, to yield a timescale for 
the starburst activity visible at ultraviolet (UV) wavelengths, 
and a relation between the star-formation rates at observed optical 
and submm wavelengths.	
For our brightest object, N2.4 (Smail et al.\ 2003) 
the best fit is for
a $\sim$5\,Myr-long instantaneous burst of star formation. The implied
star formation rate is about 1000\,M$_\odot$\,yr$^{-1}$, which 
is similar to that infered directly from its
submm and radio flux densities.
While not all sources may contain such a short burst of star-formation
dominated energy output, this example is likely to define the paradigm
for the nature of a significant fraction of SCUBA galaxies.

The morphologies of a subset of SCUBA galaxies have recently been studied 
with {\it HST}-STIS 
by Chapman et al.\ (2003d). Fig.~3 shows a montage of several representative
examples, typically revealing multi-component, distorted galaxy systems
that are reminiscent of mergers in progress.
There are no examples of isolated, compact sources, and we conclude 
generally that the morphologies of SCUBA galaxies are generally 
consistent with hierarchical galaxy
formation scenarios in which the most intense activity occurs when gas-rich, 
high-redshift galaxies collide and merge. Although they could be the sites 
of spheroid formation, most spheroids do not form in a monolithic collapse of 
primordial gas clouds at extreme redshifts.

\begin{figure}
\plottwo{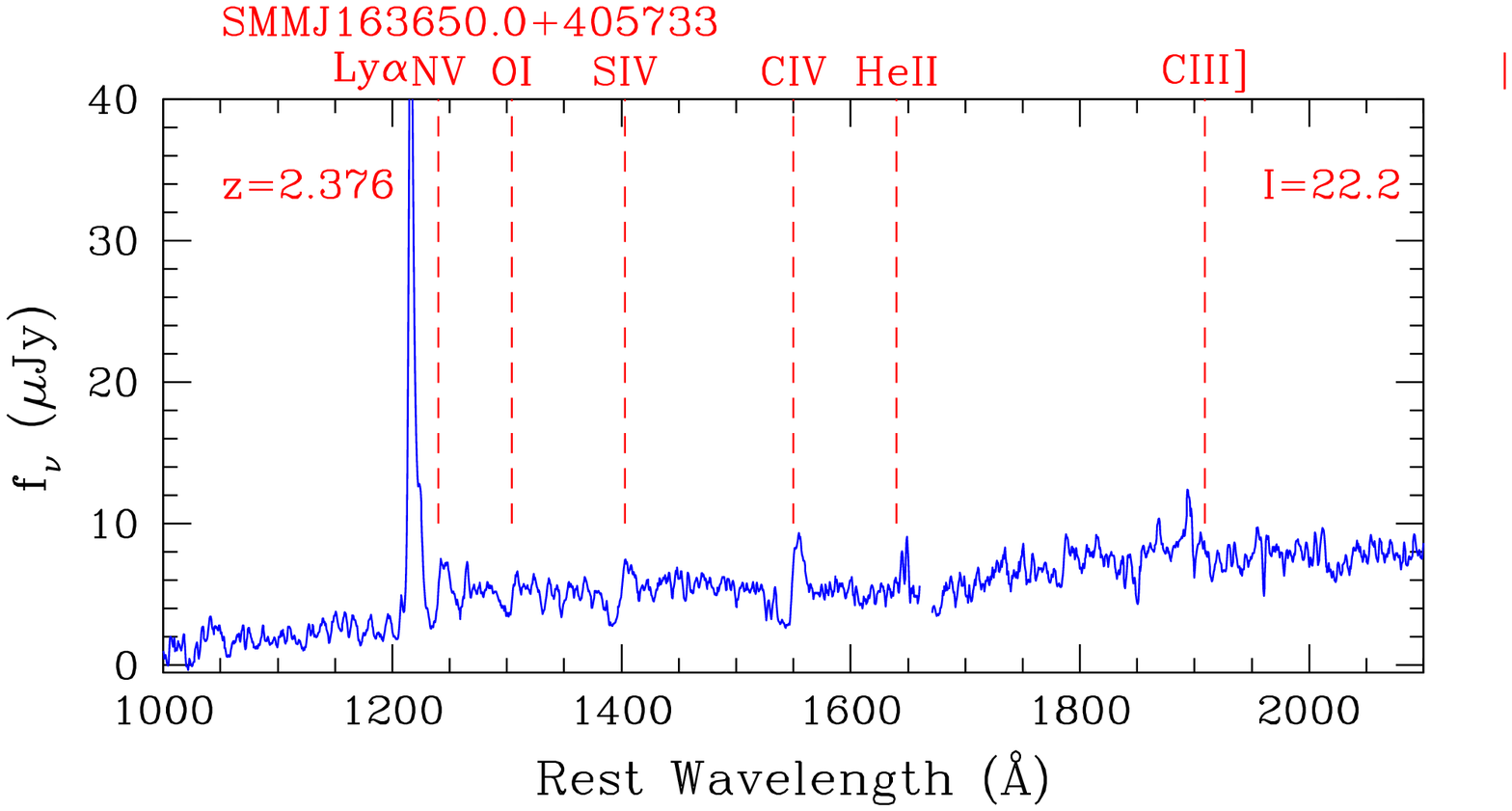}{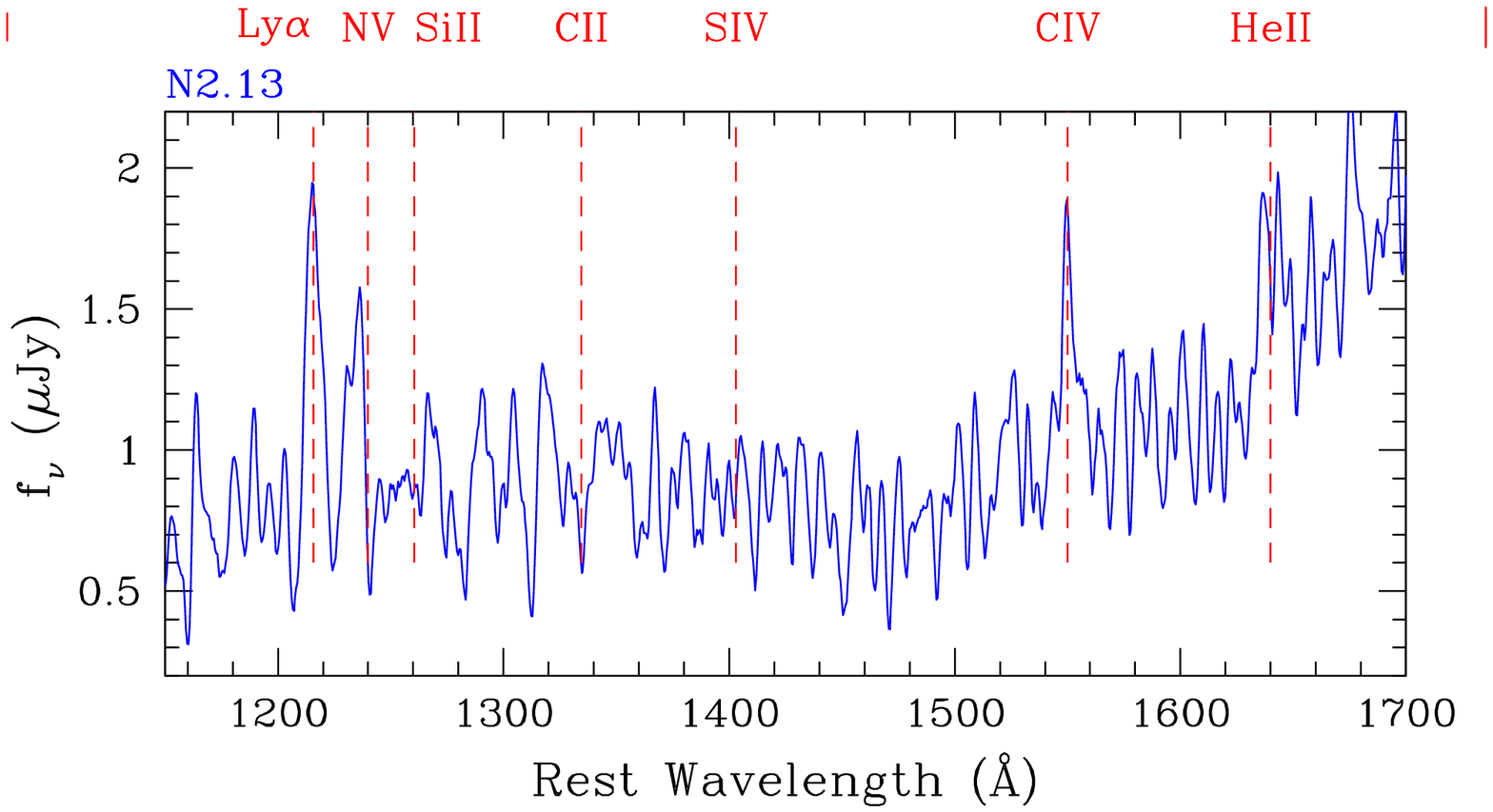}
\plottwo{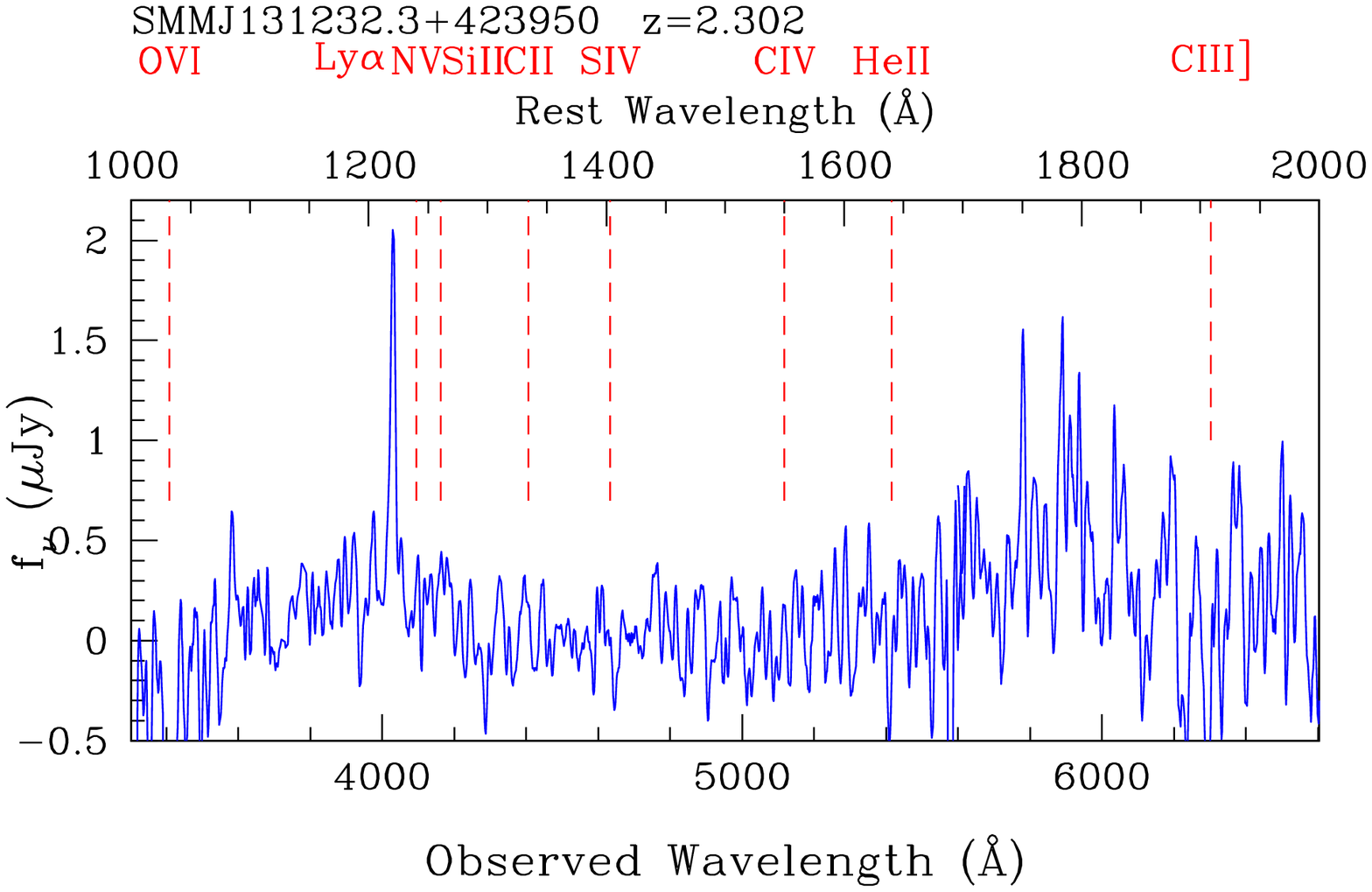}{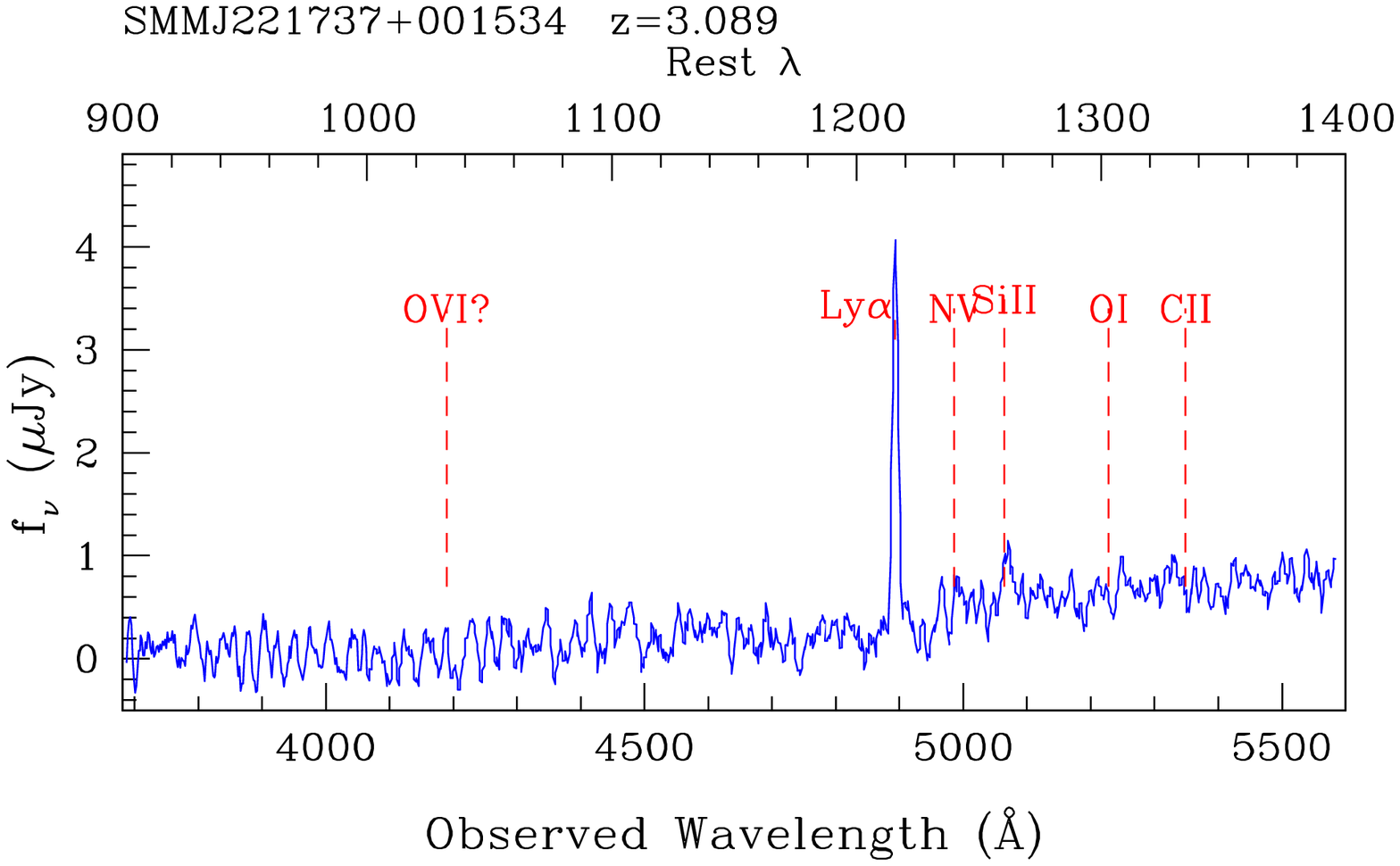}
\caption{
The rest-UV spectrum of several radio--SCUBA galaxies in our sample: 
{\bf upper left} N2.4, 
showing P-Cygni absorption line 
profiles, and a host of stellar and interstellar features. The spectrum
is best fit by a $\sim$5\,Myr-old starburst, with an inferred star-formation 
rate of 1000\,M$_\odot$\,yr$^{-1}$.
{\bf upper right} N2.13 revealing strong Ly$\alpha$ and CIV suggestive of
a narrow line AGN.
{\bf lower left} SSA13.13, showing only Ly$\alpha$, but recently detected
in H$\alpha$ using near-IR narrow-band imaging (Smail et al.\ in preparation).
{\bf lower right} SSA22.4, showing strong Ly$\alpha$ and other weaker
emission lines.
}
\end{figure}

\begin{figure}
\plotone{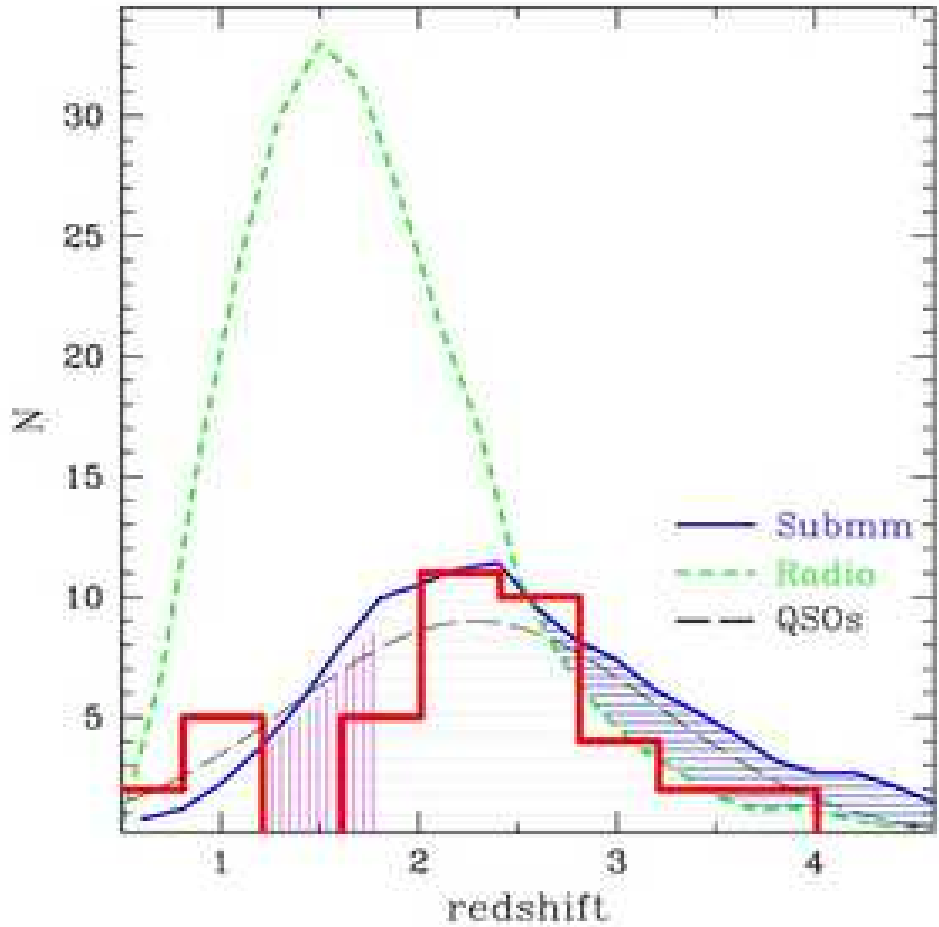}
\caption{
The observed histogram of the redshift distribution for our 55
radio-identified
SCUBA galaxies. Curves derived for
a model of the radio/submm galaxy populations are overplotted
(Chapman et al.\ 2003c, Lewis et al.\ 2003),
suggesting that the redshifts
of the sources missed in the radio identification process lie mostly 
at moderate redshifts
between the radio and submm model tracks.
A sample of radio-selected QSOs is also overplotted, revealing a remarkable
similarity with our observed distribution for submm galaxies.
}
\end{figure}
 
\begin{figure}
\plotone{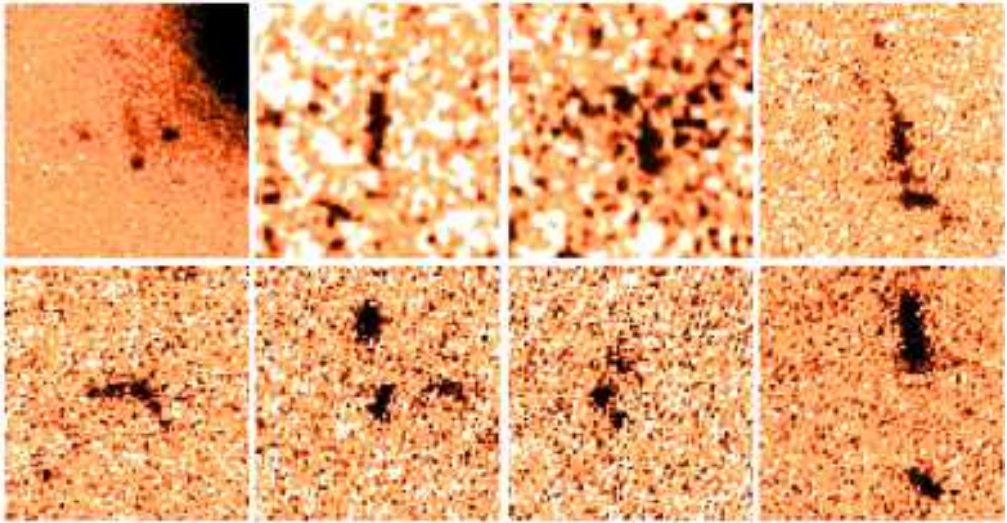}
\caption{
The morphologies of 8 SCUBA galaxies as studied with {\it HST}-STIS 
(Chapman et al.\ 2003d; 4" images
centered on the radio position). All are 
multi-component, distorted galaxy systems
reminiscent of mergers in progress.
}
\end{figure}

\section*{References}
Alexander, D., et al., ApJ, 2003, in press (astro-ph/0211267)\\
Barger, A., et al., Nature, 1998, 394, 293\\
Barger, A., Cowie, L., Richards, E., 2000, AJ 119, 2092\\
Blain, A.W., Smail, I., Ivison, Kneib, J.-P., 1999, MNRAS, 302, 632\\
Blain, A.W., Smail, I., Ivison, Kneib, J.-P., Frayer D.T., 2002, 
	Physics Reports, 369, 111 (astro-ph/0202228)\\ 
Blain, A.W., Chapman, S.C., Smail, I., Ivison, R., ApJ, submitted\\
Carilli, C., et al., 2001, ApJ, 555, 625\\
Chapman, S.C., Richards, E., Lewis, G., Wilson, G., Barger, A.,
        2001, ApJ, 548, L147\\
Chapman, S.C., Lewis, G.F., Scott, D., Borys, C., Richards, E., 2002a, ApJ,
	570, 557\\
Chapman S.C., Barger A., Cowie L., et al., 2003a, ApJ, 585, 57\\
Chapman, S.C., Blain, A., Ivison, R., Smail, I., 2003b, 
	Nature, April 17 issue\\
Chapman, S.C., Helou, G., Lewis, G., Dale, D., 2003c, ApJ, in press, astro-ph/0301233\\ 
Chapman, S.C., Windhorst, R., Odewahn, S., Haojing, Y., Muxlow, T., ApJ, 2003d, 
	submitted\\
Cowie, L., Barger, A., Kneib, J.-P., 2002, AJ, 123, 2197\\
Eales, S., et al., 1999, ApJ 515, 518\\
Guiderdoni, B., et al.\ 1998, MNRAS, 298, 708\\
Helou, P., et al., 1985, ApJ 440, 35\\ 
Hughes, D., et al., Nature, 1998, 394, 241\\
Isaak, K., et al., 2002, MNRAS, 329, 149\\
Ivison, R., et al., 1998, MNRAS, 298, 583\\ 
Ivison, R., et al., 2002, MNRAS, 337, 1\\ 
Lewis, G., Chapman, S., Helou, G.\ 2003, ApJ, submitted\\
Magorrian J. et al., 1998, AJ, 115, 2285\\
Nandra, P., et al.\ 2002, ApJ, \\
Neri, R., et al.\ 2003, A\&A, in preparation\\
Shaver, P., et al., 1998, APS Conf series 156,
Highly Redshifted Radio Lines,
ed. C.Carilli, S.Radford, K. Menten, G. Langston, (San Francisco: ASP163)\\
Smail, I., Ivison, R.J., Blain, A.W., 1997, ApJ 490, L5\\
Smail, I.,  Ivison, R.J., Blain, A.W., Kneib, J.-P., 2002, MNRAS, 331, 495\\
Smail, I., et al., 2003, MNRAS, in press, astro-ph/0303128\\
Steidel C., Adelberger K., Giavalisco M., Dickinson M.,
 Pettini M., 1999, ApJ 519, 1\\
Steidel, C., et al., 2003, ApJS, in press\\
\end{document}